 \documentclass[pmlr,twocolumn,10pt]{jmlr} 

\usepackage{float}

\usepackage{booktabs}
\usepackage{siunitx}

\usepackage[switch]{lineno}

\usepackage{enum
item}       

\usepackage{multirow}

\usepackage{xcolor,colortbl}
\usepackage{color, colortbl} 
\usepackage{nicematrix}

\newcommand{\Sref}[1]{\S\ref{#1}}

\definecolor{highlight}{rgb}{1.0,0.90,0.8}	
\definecolor{Gray}{gray}{0.96}

\definecolor{lightBlue}{rgb}{0.78, 0.85, 1.0}
\definecolor{lightOrange}{rgb}{0.88, 0.95, 1.0}
\definecolor{lightRed}{rgb}{1.0, 0.85, 0.85}
\usepackage{tcolorbox}
\usepackage{multirow}

\newtcbox{\bluebox}{on line, box align=base, colback=lightBlue,colframe=white,size=fbox,arc=3pt, before upper=\strut, top=-2pt, bottom=-4pt, left=-2pt, right=-2pt, boxrule=0pt}

\newtcbox{\orangebox}{on line, box align=base, colback=lightOrange,colframe=white,size=fbox,arc=3pt, before upper=\strut, top=-2pt, bottom=-4pt, left=-2pt, right=-2pt, boxrule=0pt}

\newtcbox{\redbox}{on line, box align=base, colback=lightRed,colframe=white,size=fbox,arc=3pt, before upper=\strut, top=-2pt, bottom=-4pt, left=-2pt, right=-2pt, boxrule=0pt}

\newtcbox{\whitebox}{on line, box align=base, colback=white,colframe=white,size=fbox,arc=3pt, before upper=\strut, top=-2pt, bottom=-4pt, left=-2pt, right=-2pt, boxrule=0pt}

\newcommand{\equal}[1]{{\hypersetup{linkcolor=black}\thanks{#1}}}

\theorembodyfont{\upshape}
\theoremheaderfont{\scshape}
\theorempostheader{:}
\theoremsep{\newline}

\jmlrvolume{259}
\jmlryear{2024}
\jmlrsubmitted{LEAVE UNSET}
\jmlrpublished{LEAVE UNSET}
\jmlrworkshop{Machine Learning for Health (ML4H) 2024} 

\title[Automating Feedback Analysis in Surgical Training]{Automating Feedback Analysis in Surgical Training: \\
Detection, Categorization, and Assessment}

\author{
    \Name{Firdavs Nasriddinov}\equal{These authors contributed equally} \quad \Name{Rafal Kocielnik}\footnotemark[1] \Email{\{firdavs,rafalko\}@caltech.edu}
\AND
    \Name{Arushi Gupta} \Email{agupta5@caltech.edu}\\
    \addr California Institute of Technology, USA
\AND
\Name{Cherine Yang
} \Email{cherine.yang@cshs.org}\\
\addr Cedars-Sinai Medical Center, USA
\AND
 \Name{Elyssa Wong} \Email{eywong@usc.edu}\\
\addr University of Southern California, USA
\AND
\Name{Anima Anandkumar} \Email{anima@caltech.edu}\\
\addr California Institute of Technology, USA
\AND
\Name{Andrew J. Hung} \Email{andrew.hung@cshs.org}\\
\addr Cedars-Sinai Medical Center, USA
}

\begin{document}

\maketitle

\begin{abstract}
This work introduces the first framework for reconstructing surgical dialogue from unstructured real-world recordings, which is crucial for characterizing teaching tasks. In surgical training, the formative verbal feedback that trainers provide to trainees during live surgeries is crucial for ensuring safety, correcting behavior immediately, and facilitating long-term skill acquisition. However, analyzing and quantifying this feedback is challenging due to its unstructured and specialized nature. Automated systems are essential to manage these complexities at scale, allowing for the creation of structured datasets that enhance feedback analysis and improve surgical education. Our framework integrates voice activity detection, speaker diarization, and automated speech recognition, with a novel enhancement that 1) removes hallucinations (non-existent utterances generated during speech recognition fueled by noise in the operating room) and 2) separates speech from trainers and trainees using few-shot voice samples. These aspects are vital for reconstructing accurate surgical dialogues and understanding the roles of operating room participants. Using data from 33 real-world surgeries, we demonstrated the system’s capability to reconstruct surgical teaching dialogues and detect feedback instances effectively (F1 score of 0.79$\pm$0.07). Moreover, our hallucination removal step improves feedback detection performance by $\approx$14\%. Evaluation on downstream clinically relevant tasks of predicting Behavioral Adjustment of trainees and classifying Technical feedback, showed performances comparable to manual annotations with F1 scores of 0.82$\pm$0.03 and 0.81$\pm$0.03 respectively. These results highlight the effectiveness of our framework in supporting clinically relevant tasks and improving over manual methods.
\end{abstract}


\paragraph*{Data and Code Availability}
Due to the nature of the data it will only be available on request. The code is publicly available on \href{https://github.com/firdavsn/SurgicalFeedbackAI}{our github}.

\paragraph*{Institutional Review Board (IRB)} The data used was collected under the IRB of the University of Southern California (HS-17-00113).

\begin{figure*}[ht!]
  \begin{center}
\includegraphics[width=1.0\textwidth]{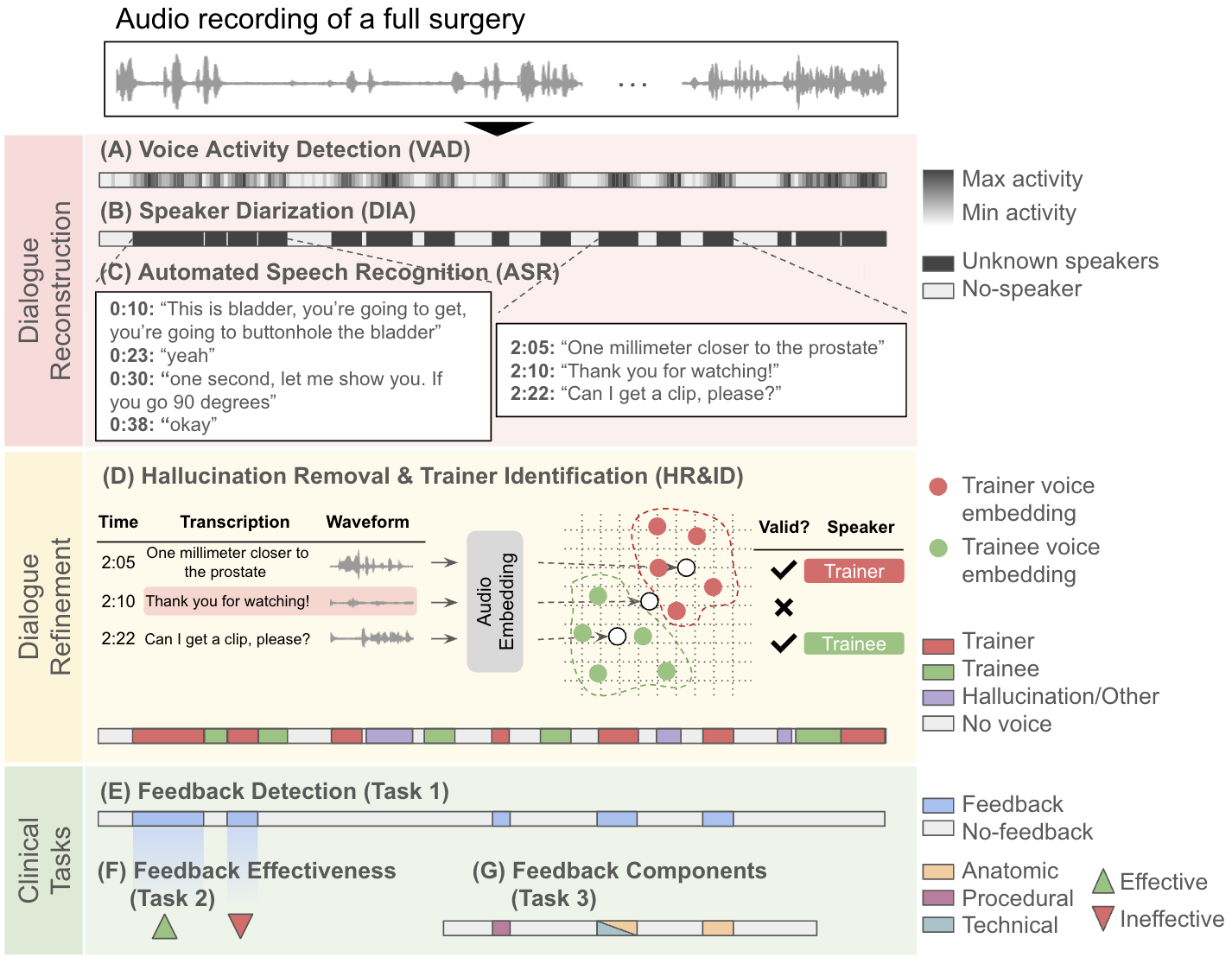}
\vspace{-20.0pt}
  \caption{Overview of our automated surgical feedback detection and assessment framework, organized into:
    Dialogue Reconstruction, which integrates (A) Voice Activity Detection (VAD) to detect timespans of speech. (B) Speaker Diarization (DIA) to differentiate speech from different speakers. (C) Automated Speech Recognition (ASR) to transcribe the audio into text.
    Dialogue Refinement focuses on (D) Hallucination Removal \& Trainer/Trainee Identification which ensures the correct identification of speakers and removes irrelevant audio snippets.
    Clinical Tasks step applied our framework to clinically relevant tasks from  \cite{wong2023development}, including (E) Feedback Detection (Task 1) which identifies when feedback is delivered. (F) Feedback Effectiveness (Task 2) which evaluates the impact and effectiveness of the feedback.
    (G) Feedback Components (Task 3) which categorizes the feedback into Anatomic, Procedural, and Technical. }
  
  \label{fig:main_steps}
  \end{center}
  \vspace{-22.0pt}
\end{figure*}

\vspace{-5.0pt}
\section{Introduction}
\label{sec:intro}

\paragraph{Importance:} 
Formative verbal feedback delivered by a trainer to a trainee during surgical procedures is crucial for immediate correction and guidance \citep{wong2023development} as well as for fostering long-term skill acquisition \citep{agha2015role}. High-quality feedback has been shown to significantly enhance intraoperative performance \citep{bonrath2015comprehensive}, accelerate surgical skill development \citep{ma2022tailored}, and bolster trainees' sense of autonomy \citep{haglund2021surgical}. The quality of communication has also been directly linked to operative outcomes \citep{d2020evaluating}. This underscores the importance of understanding current practices of feedback delivery on the quality of surgical education \citep{el2017feedback}. 
Given the critical role of feedback in surgical training and patient outcomes, automating feedback analysis is essential. It can help streamline the understanding of current practices, enhance trainer-trainee communication, and reduce inconsistencies in manual analysis. By improving training methods and creating structured datasets, automation enables the development of standardized, cost-efficient guidance systems \citep{ma2024artificial}, ultimately elevating the quality of surgical education and patient care.

\paragraph{Challenges:} 
Effectively quantifying and analyzing real-world surgical feedback at scale presents significant challenges. The contextual understanding of surgery, including its phases, specific procedures, and instructional tasks, necessitates deep domain expertise \citep{haque2024competency}. Differentiating genuine feedback from mere comments or unrelated remarks requires a thorough familiarity with operating room interactions \citep{wong2023development}. Moreover, discerning which feedback is actionable and merits detailed annotation demands surgical teaching expertise. Additionally, the often extensive durations of surgeries coupled with the complex social dynamics among medical staff lead to many exchanges that do not qualify as clinically relevant feedback. Due to these challenges, previous studies have depended on labor-intensive manual identification and annotation of feedback by trained human raters \citep{wong2023development, hauge2001reliability, blom2007analysis}. Such manual annotation requires considerable time and effort from skilled annotators and is impractical to perform at scale, thereby hindering both the systematic quantification of feedback and the creation of structured, clinically-aligned datasets essential for developing automated feedback delivery systems \citep{laca2022using}.

\paragraph{Approach:} 
We analyze full-length recordings of live surgical cases to automatically detect surgical feedback delivery moments based on the clinically validated definition of feedback from \cite{wong2023development}. We further automatically categorize the content of this feedback and assess its effectiveness. We develop a \textit{framework for reconstructing teaching interactions in the operating room} 
consisting of 3 consecutive steps as shown in Figure \ref{fig:main_steps}:

\begin{enumerate}[leftmargin=*, itemsep=0.0mm, topsep=5pt, 
]
    \item \textbf{Dialogue Reconstruction}: Combines Voice Activity Detection (VAD) to identify speech timespans in continuous audio, Speaker Diarization (DIA) to provide timespans for speech from different speakers, and Automated Speech Recognition (ASR) to transcribe audio within each timespan.

    \item \textbf{Dialogue Refinement}: Removes ASR and DIA hallucinations and identifies trainer/trainee speaking turns by leveraging cosine similarity between embeddings of candidate speech segments and few-shot embedded voice samples from the trainer and trainee present in the operating room.

    \item \textbf{Clinical Task Evaluation}: We demonstrate the crucial importance of each step on downstream clinically relevant tasks form \cite{wong2023development}: \textit{Feedback Detection},  \textit{Feedback Effectiveness Assessment}, and \textit{Feedback Components Identification}.
\end{enumerate}

We perform experiments on recordings of 33 surgeries from prior work on feedback  \citep{wong2023development}, which contains clinically validated human annotations of 4.2k feedback instances along with component and trainee behavior annotations (Table \ref{tab:dataset-stats}). 

\begin{table*}[t!]
\centering
\small
\caption{Statistics of the dataset obtained from \cite{wong2023development} containing full video and audio recordings of 33 surgeries along with available human annotations for different clinically relevant tasks which we support in our feedback detection framework.}
\vspace{-6.0pt}

\begin{tabular}{p{0.7in}lp{3.0in}rr}
\toprule
 \textbf{Task} & \textbf{Dimension} & 
 \textbf{Definition} &
 \textbf{Len/Count} & \textbf{\% Pos}\\
 \midrule

 Dialogue & - & Verbal interactions in the operating room. & $78h\: 
 52m$ & - \\

 \midrule

 \multirow{2}{0.7in}{Feedback Detection} & - & Any dialogue intended to modify trainee thinking or behavior. & 4210 & - \\
 \midrule
 
 \multirow{2}{0.7in}{Component Classification} & Anatomic & Familiarity with anatomic structures and landmarks. & 1194 & 28.4\% \\
 \addlinespace
 & Procedural & Pertains to timing and sequence of surgical steps. & 851 & 20.2\% \\
 \addlinespace
 & Technical & Performance of a discreet task with appropriate knowledge of exposure, instruments, and traction. & 3489 & 82.9\% \\

 \midrule
 \multirow{2}{0.7in}{Feedback Effectiveness} & Behavioral Adj. & Behavioral adjustment made by the trainee that corresponds directly with the preceding feedback & 1866 & 44.3\% \\
 \addlinespace
 & Verbal Ack. & Verbal or audible confirmation by the trainee confirming that they have heard the feedback & 1944 & 46.2\% \\
 \bottomrule

\end{tabular}
\vspace{-6.0pt}
\label{tab:dataset-stats}

\end{table*}

\paragraph{Findings:}
\begin{itemize}[leftmargin=*, itemsep=0.0mm, topsep=5pt]
    \item We show an ability to identify feedback utterances effectively with F1 of 0.79$\pm$0.07 (Table \ref{tab:results-feedback-detection}).
    
    \item Evaluation on the downstream clinically relevant tasks shows performance comparable to manual annotations: F1 of 0.82$\pm 0.03$ for predicting \emph{Behavioral Adjustment} of a trainee, F1 of 0.81$\pm 0.03$ for classifying \emph{Technical} feedback (Table \ref{tab:results-donwstream-tasks}). 
    
    \item We show that our Hallucination Removal step improves performance on all tasks with the most performance gains in feedback detection with an increase of $\approx$14\%.

\end{itemize}

\paragraph{Contributions:}

\begin{itemize}[leftmargin=*, itemsep=0.0mm, topsep=5pt]
    \item To the best of our knowledge, we are the first to attempt highly automated detection and assessment of surgical feedback from real-world teaching interactions in the operating room.
    \item We provide a robust evaluation of our framework on 33 real-world surgical cases involving unseen surgeries and clinically relevant downstream tasks.
    \item Aside from effectively combining existing voice tasks, we also introduce a novel step of Hallucination Removal \& Speaker Identification (Figure \ref{fig:main_steps}), which leverages audio representations to identify trainers and filter out hallucinated text transcription fragments based on audio latent space.
\end{itemize}

\vspace{-5pt}
\section{Background and Related Work}
\label{sec:background}

\paragraph{Analysis of Feedback in Operating Room.}
Historical efforts in annotating surgical teaching interactions have primarily depended on manual observation. \cite{hauge2001reliability} and \cite{blom2007analysis} developed categorization schemes through manual annotation of recorded surgeries, involving hundreds of teaching behaviors. \cite{ramprasad2024language} manually transcribed 615 minutes of operating room interactions. \cite{wong2023development} analyzed 29 surgical videos and 3,711 interactions manually, establishing a validated feedback classification and differentiating effective from ineffective feedback across various surgical stages and trainee experience levels. However, the scalability of these manual methods is limited, impacting the automation of teaching feedback systems and the generation of comprehensive, clinically relevant datasets needed for automation.

Recent efforts targeted the partial automation of feedback analysis in surgical training. \cite{kocielnik2023deep} automated the feedback categorization scheme validated by \cite{wong2023development}, using multimodal information. However, this automation was limited to the categorization phase, with the initial detection of feedback moments still dependent on manual annotations. Our work overcomes this limitation by introducing an automated system that detects feedback directly from raw recordings of surgical cases using validated definitions from \cite{wong2023development}. We enhance the system's credibility by rigorously comparing its output to human annotations in clinically relevant tasks such as feedback categorization and trainee behavior change prediction.

\paragraph{Speech Reconstruction in Surgery.}
\label{par:hall_rem}

Research on speech recognition in healthcare has primarily targeted non-surgical settings \citep{schaaf2021you, corbisiero2023speech}
, leaving significant gaps in the surgical context. The complexity of the operating room (OR) poses unique challenges for automated speech analysis systems due to non-standardized communication, background noise \citep{hasfeldt2010noise}, and the dynamic interplay of multiple roles including surgeons, nurses, and anesthesiologists \citep{blom2007analysis, gardezi2009silence}.

In this setting, the hallucinations in transcribed speech, which can misrepresent critical verbal exchanges is a key issue \citep{koenecke2024careless, kuhn2024measuring, hasfeldt2010noise}. \cite{magesh2024hallucination} reports that ASR hallucinations can affect 17\% to 33\% of content in specialized settings. Prior work tackled this through diverse methods: using multi-step verification processes \citep{taki2024mitigation}, audio-visual alignment \citep{zhang2024visual},
majority voting from multiple ASR runs \citep{koenecke2024careless}, 
or directly predicting hallucinated outputs \citep{serai2022hallucination}. In surgical settings, hallucinations manifest differently, best described as \emph{``non-existent utterances generated during speech recognition fueled by noise in the operating room''} in line with \cite{koenecke2024careless}.
These can occur across multiple steps of the dialogue reconstruction pipeline—voice activity detection, speaker diarization, and ASR transcription—often producing repetitive affirmations like ``Yeah'', ``good'', ``thank you'' rather than factual inaccuracies. Such errors are particularly problematic in the OR, where hallucinated critical verbal exchanges can be misrepresented as approvals from the trainer or trainee acknowledgments.

Our approach enhances dialogue reconstruction steps to address these challenges by filtering hallucinated responses using voice samples from the surgical team, improving both the accuracy and relevance of dialogue reconstruction in the OR. At the same time, it provides critical identification of speaking turns from trainer and trainee, which is crucial for detecting clinically relevant feedback. Our method represents a targeted solution to overcome the specific difficulties of dialogue reconstruction in surgery.

\vspace{-5pt}
\section{Methods}
\label{sec:methods}

\begin{table*}[ht]
\centering
\begin{tabular}{p{2.3cm}ll lcc}
\textbf{Technique} & \textbf{Data Processing} & \textbf{Classifier} & \textbf{F1-bin} & \textbf{Precision} & \textbf{Recall} \\
\hline

\multirow{4}{2.3cm}{Fixed-Window (baseline)}
& Voice Activity Detection (VAD) & - & 0.42$_{\pm0.20}$  & 0.28$_{\pm0.17}$   & 1.00$_{\pm0.00}$ \\
& + Audio & Wav2Vec2  & 0.52$_{\pm0.11}$ 
& 0.49$_{\pm0.19}$   & 0.59$_{\pm0.11}$ \\
& + Text (ASR) & BERT & 0.59$_{\pm0.13}$ 
& 0.55$_{\pm0.18}$   & 0.66$_{\pm0.10}$ \\
& + Text (ASR) & GPT-4o & 0.60$_{\pm0.11}$ 
& 0.60$_{\pm0.16}$   & 0.62$_{\pm0.06}$ \\
& + Audio + Text (ASR) & Multimodal & 0.58$_{\pm0.13}$  
& 0.53$_{\pm0.19}$   & 0.67$_{\pm0.07}$ \\
\hline

\multirow{3}{2.3cm}{Dialogue Reconstruction} & Dialogue$^\mp$ & GPT-4o & 0.58$_{\pm0.11}$ 
& 0.64$_{\pm0.15}$ & 0.55$_{\pm0.09}$ \\
& + Hallucination Rem. (baseline) & GPT-4o & 0.59$_{\pm0.07}$  
& 0.57$_{\pm0.07}$ & 0.61$_{\pm0.08}$ \\
& + Hallucination Rem. \textbf{(our)} & GPT-4o & 0.66$_{\pm0.18}^{*}$  
& 0.65$_{\pm0.22}$ & 0.71$_{\pm0.11}$ \\
& + Trainer/Trainee ID$\,$ \textbf{(our)} & GPT-4o & \textbf{0.79$_{\pm0.07}^{*}$} 
& \textbf{0.76$_{\pm0.12}$} & \textbf{0.85$_{\pm0.09}$} \\
\hline
\end{tabular}
\caption{Performance on \textbf{Feedback Detection}. Fixed-Window technique classifies on rolling 10-sec audio fragments through Temporal Event Detection. Acronyms: Voice Activity Detection (VAD) - used to detect audio containing speech. Automated Speech Recognition (ASR) - used to transcribe the audio of speech to text. Speaker Diarization (DIA) - used to separate the speech from different speakers as separate timespans. $^{\mp}$Dialogue is a combination of VAD, DIA, and ASR applied in sequence and grouped with the context of past utterances leading to feedback. Statistically significant gain compared to prior step in dialogue reconstruction at $^{*}$p$<$0.05, $^{\dag}$p$<$0.1.}
\vspace{-10.0pt}
\label{tab:results-feedback-detection}
\end{table*}

\begin{table*}[ht]
\centering
\begin{tabular}{l | ll |  lll}
\textbf{Data Processing}  & \multicolumn{2}{c}{\textbf{Feedback Effectiveness}} & \multicolumn{3}{c}{\textbf{Feedback Components}} \\
                & \textbf{Beh. Adj.} & \textbf{Verb. Ack.} & \textbf{Anatomic} & \textbf{Procedural} & \textbf{Technical} \\
\hline
\rowcolor{Gray}Manual annotations & 0.78$_{\pm0.03}$ & 0.63$_{\pm0.04}$ & 0.64$_{\pm0.11}$ & 0.46$_{\pm0.19}$ & 0.78$_{\pm0.03}$ \\
Dialogue  & 0.80$_{\pm0.02}$ 
& 0.61$_{\pm0.09}$ 
& \textbf{0.69$_{\pm0.09}$} 
& 0.45$_{\pm0.18}$ 
& 0.77$_{\pm0.03}$ 
\\
$^{+}$ Hallucination Rem. & \textbf{0.82$_{\pm0.03}^{\dag}$} 
& \textbf{0.66$_{\pm0.06}^{*}$} 
& 0.65$_{\pm0.09}$ 
& \textbf{0.49$_{\pm0.17}^{\dag}$} 
& \textbf{0.81$_{\pm0.03}^{*}$} 
\\
$^{+}$ Trainer/Trainee ID & \textbf{0.82$_{\pm0.08}$} 
& 0.64$_{\pm0.04}$ 
& 0.66$_{\pm0.08}$ 
& 0.46$_{\pm0.17}$ 
& \textbf{0.81$_{\pm0.03}$} 
\\
\hline

\hline
\end{tabular}
\caption{
Performance on \textbf{Feedback Effectiveness Assessment} and \textbf{Feedback Component Classification} downstream clinically validated tasks (F1 binary). Manual annotation represents the human baseline provided in \cite{wong2023development}. Human annotation involved ground truth labels for the tasks as well as limited transcription of the feedback itself, which was used as input for prediction. This manual transcription does not include dialogue leading to feedback and often only selected phrases in trainer feedback are transcribed. Automated transcription offers a more comprehensive transcript with dialogue context, which is also used as input for the tasks. Statistically significant gain compared to prior step in dialogue reconstruction at $^{*}$p$<$0.05, $^{\dag}$p$<$0.1.
}

\label{tab:results-donwstream-tasks}
\vspace{-14.0pt}
\end{table*}

\subsection{Data Acquisition}
\label{sec:data_acq}
Our study utilized a dataset of genuine feedback from trainers to trainees captured during real-world robot-assisted surgical procedures obtained by \cite{wong2023development} and detailed in Table \ref{tab:dataset-stats}. The dataset covers multi-organ surgical contexts across 7 procedures, involving 4 trainers and 11 trainees. It has been rigorously annotated by 3 trained human raters. Feedback was recorded using wireless microphones on the surgeons and a video recorder capturing the surgeon's point-of-view through an endoscope camera, with all data synchronously captured using an external device and the da Vinci Xi surgical robot system \citep{dimaio2011vinci}. Each instance of feedback, totaling 4210, was timestamped and manually transcribed from the audio data.

\vspace{-5pt}
\subsection{Surgical Feedback Definition}
Following the clinically validated definition by \cite{wong2023development}, surgical feedback is \emph{any dialogue intended to modify trainee thinking or behavior during a live surgery}. This feedback must be delivered in real-time by a trainer to a trainee who is actively operating the robotic console, allowing for immediate application and adjustment. The communication aims to influence the trainee’s actions, decision-making, or understanding of the surgical task at hand and must be contextually relevant to the ongoing procedure. Social conversations and unrelated discussions within the operating room are not considered feedback. 

\vspace{-5pt}
\subsection{Surgical Dialogue Reconstruction}

The process of reconstructing surgical dialogues is outlined in Figure \ref{fig:main_steps}. Initially, the entire audio stream is divided into 3-minute chunks for individual processing. Each chunk is first processed using \emph{Voice Activity Detection (VAD)} through python \emph{py-webrtcvad} module \citep{wisemanp31:online}, which assigns a value from 0 to 1 for every 10ms frame, where 0 indicates no activity and 1 maximum activity.

In addition to VAD, the raw audio undergoes \emph{Speaker Diarization (DIA)} utilizing the speaker-diarization-3.1 model from \emph{Pyannote} \citep{Bredin23DIA, Plaquet23DIA}. This method identifies segments of audio containing speech and assigns random speaker IDs such as "SPEAKER 0." Each segment detected by DIA is cross-verified with the VAD output, and segments without significant VAD activity (below a threshold of 0.3) are discarded. This threshold was determined through empirical testing of various values (see Appendix \ref{apd:vad_threshold}).

The remaining segments are then transcribed using \emph{Automated Speech Recognition (ASR)} with the \emph{Whisper-1} model \citep{radford2023robust}, pre-trained on 680k hours of labeled English speech data for accurate transcription. This model was specifically fine-tuned for speech recognition tasks. Speech data was annotated using large-scale weak supervision. 

\vspace{-5pt}
\subsection{Hallucination Removal \& Trainer / Trainee Identification} 
The output of prior steps produces significant hallucinations due to background noise in the OR. Furthermore, the lack of knowledge about the role of the speaker hinders the detection of clinically meaningful guidance. To address both issues we introduced a custom dialogue refinement step.

We manually select and annotate a set of anchor audio segments for each trainer and trainee. These anchors are chosen to represent clear instances of each individual's voice. We identified at least 5 anchors per person, balancing the need for sufficient representation with practical constraints. The selection process involves visualizing speaking (or voice activity) times and choosing diverse segments across the surgery duration. We prioritize segments with minimal background noise and clear speech. The number of anchors (5+) was determined through empirical testing (Appendix \ref{apd:cos_sim_examples} shows sufficient dissimilarity between separate speaker voices with the use of 5 anchors), to capture voice variations while remaining manageable in the applied context. These anchor segments are then embedded using \emph{Pyannote}'s ``embedding'' model \citep{Bredin2020embedding, Coria2020embedding} that is based on the x-vector TDNN-based architecture \citep{snyder2018xvectors} with 4.2M parameters.

Our method for hallucination removal and speaker role identification (Figure \ref{fig:main_steps}) processes each audio segment from the diarization step. We embed these segments using the same model as the anchors. For each segment, we compare its embedding to all anchor embeddings of the corresponding trainer and trainee using cosine similarity. We then calculate the average similarity for both trainer and trainee. 

To identify hallucinations, we apply a threshold of 0.2 to both average similarities (see Appendix \ref{apd:cossim_thresh} for determining threshold). Segments falling below this threshold for both trainer and trainee are classified as hallucinations or other speakers and excluded. Otherwise, we assign the segment to either trainer or trainee based on the higher similarity score. This approach not only removes hallucinations,  but also leverages domain knowledge to identify the critical roles of trainer and trainee, which are essential for understanding the surgical teaching context. 

\vspace{-7pt}
\section{Experiments}
\label{sec:experiments}

We evaluate our method on several clinically relevant downstream tasks: \textit{Feedback Detection (Task 1)}, \textit{Feedback Effectiveness Assessment (Task 2)}, and \textit{Feedback Component Classification (Task 3)}. We compare the performance of our approach to human annotator baselines as well as to other solutions.

We evaluate models on a test set including five unseen surgery cases covering feedback from different trainer-trainee pairs under different procedures. This test set is also representative of the whole dataset in terms of class distributions under all the tasks (see Table \ref{tab:test_set_stats} in Appendix \ref{apd:test_set_stats}). We compute binary recall, precision, and F1 score \citep{34Metric4:online} between true and predicted labels for each task.

We further statistically compare the impact of crucial \emph{Hallucination Removal} and \emph{Trainer/Trainee Identification} steps to the dialogue reconstruction without these steps using McNemar's non-parametric statistical test \citep{dietterich1998approximate} further adapted to deep learning setups by \cite{vanwinckelen2012estimating}. We use a Python implementation of McNemar's test provided in \cite{raschkas_2018_mlxtend}.

\vspace{-5pt}
\subsection{Task 1: Feedback Detection}
This task involves identifying instances where the trainer provides feedback to the trainee operating the surgical console. Performance is evaluated as correctly labeling audio segments as feedback or not, compared to human annotations. Detecting feedback occurrences is crucial for assessing feedback quality. The annotation counts are in Table \ref{tab:dataset-stats} under the ``Feedback Detection'' task. We introduce several baselines.

\subsubsection{Baselines}

\paragraph{Voice Activity Detection:}
This approach simply detects any speech and separates it from the non-speech sounds in the OR. We then use these speech times as predicted feedback instances and align them with the human expert annotations. The performance of this baseline indicates how much of the verbal interaction in the OR is related to clinically valid teaching feedback.

\paragraph{Fixed-Window Temporal Event Detection:} 

We use a 10-sec moving window with 5-sec overlap, chosen based on average feedback length. VAD is applied as a first preprocessing step, which identifies speech with sub-second precision.  For detected speech fragments, we apply \textit{Automated Speech Recognition (ASR)} with \textit{Whisper-1} \citep{radford2023robust} for text transcription. We then classify each window for feedback using Audio, Text, and Audio+Text late fusion models. We fine-tune Wav2Vec base \citep{baevski2020wav2vec} for audio classification, and BERT base \citep{devlin2018bert} for text classification. For multimodal classification, we use late fusion, extracting 256-dimension vectors from audio and text, concatenating them, and passing through fully connected layers with ReLu activation and dropout. 

For model training, we preprocess data using 10-second audio fragments (3 seconds before to account for any imperfections in annotating and 7 seconds after to capture feedback delivery). 
Audio is downsampled to 16kHz mono. 
We fine-tune each classifier (audio, text, audio+text) with 5 different IID dataset splits using a balanced set of feedback/no-feedback instances from each surgery case for training (due to class imbalance), which was further split into 80\%/20\% train/val sets, separate from the test set. All hyperparameters were selected based on performance on the validation split. Models are trained for 20 epochs with an initial learning rate of 5e-5, Adam optimizer, and linear LR reduction. The best model is determined by the highest binary F1 score. We use standard fine-tuning with all weights being trainable.

\paragraph{Hallucination Removal - Multiple ASR Runs:}
Several existing methods for ASR hallucination removal described in \Sref{par:hall_rem} address different types of hallucinations or rely on the presence of additional information not available in our context (e.g., aligned video modality, annotated hallucinations for direct prediction). One method we could compare to was proposed in \cite{koenecke2024careless} and relies on running the ASR more than once and removing the transcriptions that differ as hallucinations.

\subsubsection{Evaluation Setup}
To assess the effectiveness of our dialogue reconstruction approach for feedback detection, we run our framework to obtain audio segments with transcriptions and speaker roles. GPT-4o classifies these for feedback presence (see prompt used in Appendix \ref{apd:fb_det_prompt}), aligned with human annotations. We then apply a 5-second tolerance where we prompt GPT-4o to check if the phrase from the extracted dialogue and the human annotations have overlaps in transcriptions. This step is only needed for the alignment of ASR with human-transcribed instances for the purpose of evaluation. It is specifically relevant for situations where the beginning of the human-annotated transcription precedes the ASR transcription segment, but part of the content of the human transcription is still within the ASR transcribed content.

\vspace{-5pt}
\subsection{Task 2: Feedback Effectiveness}
This task evaluates how effectively the delivered feedback impacts subsequent trainee behavior. This is measured by human expert annotated observed behaviors of the trainee post feedback delivery in two categories of \emph{Behavioral Adjustment} - \textit{``Behavioral adjustment made by the trainee that corresponds directly with the preceding feedback''} and \emph{Verbal Acknowledgment} - \textit{``Verbal or audible confirmation from the trainee confirming that they heard the feedback''}.

\subsubsection{Baselines}
\paragraph{Human Selective Transcription:} 
We leverage text transcription from human annotators available in our dataset with the same GPT-4o classifier using the same prompt but adapted for single phrases instead of a dialogue (see Appendix \ref{apd:fb_eff_hum_annots_prompt}). Human annotators provided transcriptions of only the trainer's feedback, without any conversational context.
The details of the annotation and transcription process can be found in \cite{wong2023development}.

\subsubsection{Evaluation Setup}
We prompt GPT-4o to predict Verbal Acknowledgement and Behavioral Change (see Appendix \ref{apd:fb_eff_prompt}).
This evaluation is done on true positive feedback phrases obtained from Task 1. True labels come from aligned human annotations.

\vspace{-5pt}
\subsection{Task 3: Feedback Components}
This task requires categorizing feedback into 3 clinically validated components based on \cite{wong2023development}. 
These components represent feedback categorized as \textit{Anatomic} -  \emph{``familiarity with anatomic structures and landmarks''}, \textit{Procedural} - \emph{``timing and sequence of surgical steps''}, and \textit{Technical} - \emph{``performance of a discrete task with appropriate knowledge of factors including exposure, instruments, and traction''}. This categorization is not mutually exclusive, as feedback instances can pertain to anatomic, procedural, and technical aspects simultaneously. The annotation counts for this task can be found in Table \ref{tab:dataset-stats} under the component classification task.

\subsubsection{Baselines}
\paragraph{Human Selective Transcription:} 
We leverage the same selective transcriptions as for Task 2 (See Appendix \ref{apd:fb_clf_hum_annotprompt} for the GPT-4o prompt used in this task).

\subsubsection{Evaluation Setup}
We prompt GPT-4o to categorize feedback into Anatomic, Procedural, and/or Technical components in a multi-label fashion using the prompt in Appendix \ref{apd:fb_comp_prompt}. True labels come from aligned human annotations.

\vspace{-7pt}
\section{Results}
\label{sec:results}

We present results for Feedback Detection in Table \ref{tab:results-feedback-detection} and for other downstream tasks of Feedback Effectiveness and Feedback Components in Table \ref{tab:results-donwstream-tasks}. 

\vspace{-5pt}
\subsection{Task 1: Feedback Detection}
Our results in Table \ref{tab:results-feedback-detection} show the Precision, Recall, and F1 \citep{34Metric4:online} scores for detecting feedback using two techniques: Fixed-Window (baseline), and Dialogue Reconstruction. For the fixed-window, we see that the naive VAD-only classifier performed the worst with an F1 of 0.42$_{\pm 0.20}$, indicating that not all speech in the operating room corresponds to feedback. Further, we see that the text and multimodal classifiers perform very similarly with F1's of 0.59$_{\pm 0.13}$ and 0.58$_{\pm 0.13}$, respectively, and outperform the audio classifier that achieves an F1 of 0.52$_{\pm 0.11}$.

The results for the Dialogue row are obtained using initial dialogue reconstruction, and we see performance on par with best fixed-window classifiers. Our Hallucination Removal approach offers a big boost in performance, with an F1 of 0.66$_{\pm 0.18}$, which is more effective than the baseline Hallucination Removal approach from \cite{koenecke2024careless} with an F1 of 0.59$_{\pm 0.07}$. Finally, identifying between Trainer and Trainee for each dialogue phrase achieves the highest F1 at 0.79$_{\pm 0.07}$. Appendix \ref{apd:fb_confusion_matrices} provides confusion matrices for various ablations of feedback detection. 

High recall is more important than high precision, as it's more important to correctly detect all the feedback instances than to have the model be correct each time it determines a segment to be feedback.  This can be corrected in a subsequent human verification step. Our method performed the best with a recall of 0.71$_{\pm 0.11}$ after removing hallucinations and 0.85$_{\pm 0.09}$ after identifying between Trainer and Trainee.

\vspace{-5pt}
\subsection{Task 2: Feedback Effectiveness}
Table \ref{tab:results-donwstream-tasks} shows results for predicting \emph{Behavioral Adjustment} and \emph{Verbal Acknowledgement} under Feedback Effectiveness. For this task, the baseline is the classifications done on the manual annotations that yield an F1 of 0.78$_{\pm 0.03}$ and 0.63$_{0.04}$ for \emph{Beh. Adj.} and \emph{Verb. Ack}, respectively. We see that the results for the initial dialogue reconstruction slightly improve for \emph{Beh. Adj.} and slightly worsen for \emph{Verb. Ack.}. Further, Hallucination Removal improves both metrics with F1 of 0.82$_{\pm 0.03}$ for \emph{Beh. Adj.} and 0.66$_{\pm 0.06}$ for \emph{Verb. Ack}. Finally, identifying Trainer/Trainee does not help further.

\vspace{-5pt}
\subsection{Task 3: Feedback Components}
Table \ref{tab:results-donwstream-tasks} also shows results for classifying feedback as \emph{Anatomic}, \emph{Procedural}, and/or \emph{Technical} under Feedback Components. Similar to Feedback Effectiveness, the baseline leverages  manual annotations in classification. For classifying \emph{Anatomic} feedback, the base Dialogue performs best at an F1 of 0.69$_{\pm 0.09}$ while Hallucination Removal and Trainer/Trainee ID still outperform the manual annotations. For classifying \emph{Procedural} feedback, Hallucination Removal yields the best results at 0.49$_{0.17}$ with Trainer/Trainee ID performing on par with manual annotations. For classifying \emph{Technical} feedback, we see that Hallucination Removal and Trainer/Trainee ID are equal and the highest with an F1 of 0.81$_{\pm 0.03}$ for both. 

\vspace{-7pt}
\section{Discussion}
Our framework achieved F1 scores of 0.79$_{\pm 0.07}$ in feedback detection and up to 0.82$_{\pm 0.08}$ in analyzing effectiveness, demonstrating the high feasibility of automated surgical training analysis. We enable highly automated quantification and analysis of feedback in real surgeries, with implications for improving training practices and patient care. Given the wide coverage of procedures, tasks, anatomic settings, as well as trainer-trainee interactions in our dataset as in \Sref{sec:data_acq}, our approach is likely to generalize to various educational clinical settings where guidance is performed via verbal feedback and dialogue. Our approach relies on data organization and preprocessing, rather than on a specific fine-tuned model, which should further reduce the reliance on a particular dataset.

\paragraph{Surpassing Human Annotation in Downstream Tasks}

Our automation surpassed human annotations in several downstream tasks, primarily because it captured more comprehensive contextual information. Unlike humans, who transcribed only feedback itself, our method transcribed the entire dialogue involving rounds of discussion leading to feedback,
better reflecting the true nature of interactions in the operating room \citep{wong2023development}. This full transcription allowed for a more effective grouping of feedback instances, combining several human-labeled segments into a single, more informative one. Secondly, some of the human transcriptions used abbreviations and selectively transcribed parts of what actually has been said. This is due to the manual and cognitive effort associated with literal transcription, which further motivates the benefits of automation. Finally, the automated system applied consistent classification criteria across all samples, avoiding the biases and inconsistencies often introduced by human annotators \citep{kiyasseh2023multi}.

\paragraph{Use of Few-shot Speech Samples}

Our dialogue refinement step using speech samples from known speakers enhances feedback detection but requires collecting clean speech before surgery and assumes consistent vocal traits during the actual procedure. This approach is effective for a fixed group of speakers. For unknown speakers or uncollectable samples, options like unsupervised speaker diarization \citep{xylogiannis2024multisensory} or role-based speaker recognition \citep{bellagha2020speaker} offer alternatives, dependent on the surgical setting’s constraints. Although our method improves accuracy, it demands thorough consideration of these implementation factors.

\paragraph{Limitations and Future Directions}
Our research identifies key areas for future exploration. First, enhancing feedback analysis through the integration of visual data from surgeries could link verbal feedback to specific surgical actions, enriching the context significantly \citep{kocielnik2023deep}. Second, adapting our methods for real-time feedback during surgeries would not only improve teaching assessments but also help in documenting educational elements in a live setting \citep{akbari2023development}. Lastly, investigating how feedback patterns evolve over time and developing methodologies to track and improve feedback delivery could provide deeper insights into the effectiveness of surgical training and pedagogical evolution \citep{ma2021innovations}.

\vspace{-7pt}
\section{Conclusion}
This work introduced a novel automated method for feedback detection in surgical education, utilizing dialogue reconstruction, hallucination removal, and speaker identification. Our method has shown robust performance, even surpassing human annotation. This scalable system promises to improve educational strategies and patient outcomes, marking a significant advancement in the automation of surgical education analysis. It sets the stage for future developments in real-time analysis and automated feedback delivery systems, with broad implications for healthcare training and patient care.

\vspace{-6pt}
\acks{Research supported by NCI NIH under Award Number R01CA251579, R01CA273031. Content is solely the responsibility of authors and does not necessarily represent official views of NIH.}

\bibliography{references}

\appendix

\newpage
.
\section{Automated and  Manual Feedback Alignment Example}
\label{apd:aligned-feedback-example}

\begin{table}[H]
\small
\centering
\begin{tabular}{p{5.5cm}p{3.5cm}p{1cm}p{3.5cm}p{1cm}}
\toprule
\textbf{Context} & \textbf{Phrase} & \textbf{Pred Label} & \textbf{Aligned Human Annotations} & \textbf{True Label} \\
\midrule
    \textbf{00:37:58 Trainer}: "That's good, I like that.", 
    
    \textbf{00:38:12 Trainer}: "Make sure you stay in the same plane, more or less R.",
    
    \textbf{00:38:20 Trainer}: "And about there you can start doing a little mild smile towards the prostate.",
    
    \textbf{00:39:11 Trainer}: "So open up wide. This is what I mean by not digging yourself in the hole. Okay, I want all the lateral stuff opened up or dropped.",
    
    \textbf{00:39:25 Trainer}: "The reason for that is it gives you a sense of the prostate's contour as opposed to this distorted thing where you have...",

& \textbf{00:39:44 Trainer}: "So even more than what you've actually done, I would have just buzzed that right there when you can see it, whatever the, yeah, butternut thing, uh-huh." & TRUE & 
    \textbf{00:39:44}: "so even more than what you've actually done... I would've...",
    
    \textbf{00:39:48}: "just buzz that right there when you can see it... whatever the... yeah bladder neck thing"
& TRUE \\
\hline

    \textbf{00:38:12 Trainer}: "Make sure you stay in the same plane, more or less R.",
    
    \textbf{00:38:20 Trainer}: "And about there you can start doing a little mild smile towards the prostate.",
    
    \textbf{00:39:11 Trainer}: "So open up wide. This is what I mean by not digging yourself in the hole. Okay, I want all the lateral stuff opened up or dropped.",
    
    \textbf{00:39:25 Trainer}: "The reason for that is it gives you a sense of the prostate's contour as opposed to this distorted thing where you have...",
    
    \textbf{00:39:44 Trainer}: "So even more than what you've actually done, I would have just buzzed that right there when you can see it, whatever the, yeah, butternut thing, uh-huh.",
    
& \textbf{00:40:14 Trainee}: "Yeah. Sorry, I wasn't aware of that. No problem." & FALSE & & FALSE \\
\hline

\end{tabular}
\label{table:feedback}
\end{table}

\newpage
.
\newpage
\section{VAD Threshold Experiment}\label{apd:vad_threshold}

\begin{table}[H]
\centering
\begin{tabular}{l|ccc}
\textbf{VAD Thresh} & \textbf{Precision} & \textbf{Recall} & \textbf{F1-bin} \\
\hline

0 & 0.32$_{\pm 0.17}$ & 0.67$_{\pm 0.09}$ & 0.41$_{\pm 0.16}$ \\
0.1 & 0.37$_{\pm 0.18}$ & 0.67$_{\pm 0.09}$ & 0.45$_{\pm 0.15}$ \\
0.3 & 0.43$_{\pm 0.17}$ & 0.66$_{\pm 0.10}$ & \textbf{0.50$_{\pm 0.13}$} \\
0.5 & 0.39$_{\pm 0.18}$ & 0.65$_{\pm 0.07}$ & 0.49$_{\pm 0.12}$ \\
\hline

\end{tabular}
\label{tab:vad_threshold_tunning}
\caption{Fixed-length temporal event detection metrics for varying VAD thresholds averaged across results on validation set using text binary classifier models.}
\end{table}

\section{Label Distributions for 5 Unseen Test Set Surgeries }\label{apd:test_set_stats}

Table \ref{tab:test_set_stats} summarizes the class distribution for the test set of 5 unseen surgeries across all the tasks. The test set has been chosen to cover feedback from different trainer-trainee pairs under different procedures and represents 22.8\% of the data (961 of 4210 total instances). The label distributions (\%) are representative of the whole dataset (see Table \ref{tab:dataset-stats}).

\begin{table*}[h]
\centering
\begin{tabular}{llcc}
\hline
\textbf{Task} & \textbf{Dimension} & \textbf{Count} & \textbf{\% Pos} \\ \hline
Feedback Detection & Instances & 961 & \\ \hline
\multirow{3}{*}{Component Classification} 
& Anatomic & 290 & 30.2\% \\
& Procedural & 194 & 20.2\% \\
& Technical & 734 & 76.4\% \\ \hline
\multirow{2}{*}{Feedback Effectiveness}   
& Behavioral Adjustment & 517 & 53.8\% \\ 
& Verbal Acknowledgment & 415 & 43.2\% \\ \hline
\end{tabular}
\caption{Summary of label distribution for the test set of 5 unseen surgeries across different tasks.}
\label{tab:test_set_stats}
\end{table*}

\section{GPT-4o Feedback Detection Prompts}\label{apd:fb_det_prompt}

\paragraph{System Prompt:} You are a binary classifier that determines whether a given phrase contains delivery of feedback from a trainer to a trainee where the trainee is conducting urology surgery using the da Vinci robot. The dialogue is between two speakers, a trainer and a trainee. There are multiple turns in the dialogue where the same speaker can go back to back because a piece of dialogue from the other speaker might not have been picked up or because the other speaker didn't speak as much (usually the trainer speaks more than the trainee). There can be 6 types of feedback:
                 
1. Anatomic: familiarity with anatomic structures and landmarks. i.e. 'Stay in the correct plane, between the 2 fascial layers.'

2. Procedure: pertains to timing and sequence of surgical steps. i.e. 'You can switch to the left side now.'

3. Technical: performance of a discrete task with appropriate knowledge of factors including exposure, instruments, and traction. i.e. 'Buzz it.'

4. Praise: a positive remark. i.e. 'Good job.'

5. Criticism: a negative remark. i.e. 'It should never be like this.'

\paragraph{User Prompt:} 
Classify whether the following phrase contains the delivery of feedback considering the given context of the last couple turns in the dialogue where the phrase is the last entry in the context.

Format your response as follows. DO NOT DO ANY OTHER FORMATTING.:

\{'feedback': 'yes'\} if the dialogue contains feedback

\{'feedback': 'no'\} if the dialogue does not contain feedback

Context:

$\langle$ context string $\rangle$

Phrase:

$\langle$ phrase string $\rangle$

For example:

\{'feedback': 'yes'\}

\section{GPT-4o Feedback Prediction Alignment Prompt}\label{apd:fb_ali_prompt}

\paragraph{System Prompt:}
You are a binary classifier that determines whether two strings have any alignment or not. An alignment means that the two strings might have some common words or phrases that align with each other in terms of their order and/or meaning.

\paragraph{User Prompt:}
Classify whether the following two strings have any alignment or not.

Format your response as follows. DO NOT DO ANY OTHER FORMATTING.:

\{'alignment': 'yes'\} if the two strings have any alignment

\{'alignment': 'no'\} if the two strings do not have any alignment

For example:
\{
    'alignment': 'yes'
\}

String 1:

$\langle$ phrase $\rangle$

String 2:

$\langle$ human annotation $\rangle$

\section{GPT-4o Feedback Effectiveness Prediction From Auto Dialogue Prompt}\label{apd:fb_eff_prompt}

\paragraph{System Prompt:}
You are an AI assistant specializing in predicting trainee responses during urology surgery training using the da Vinci robot. Your task is to analyze dialogue between a trainer and a trainee, focusing on the trainee's reactions to feedback. The dialogue may contain multiple consecutive turns by the same speaker due to missed responses or varying speech patterns.

You will categorize potential trainee responses into two types:

1. Verbal Acknowledgement: This includes any verbal or audible confirmation from the trainee indicating they have heard and understood the feedback. Examples include:
   - "Okay, I see"
   - "Uh-huh, got it"
   - "Understood"
   - "Yes, I'll do that"

2. Behavioral Change: This refers to any physical or observable adjustment made by the trainee that directly corresponds to the feedback received. For example:
   - If the trainer suggests tightening a suture, the trainee immediately pulls the suture thread more tightly.

Your role is to predict which type(s) of response the trainee is likely to give based on the specific feedback provided by the trainer. Consider the context of the surgical procedure and the nature of the feedback when making your predictions.

\paragraph{User Prompt:}
Classify whether the following feedback phrase will lead to a trainee response, where a trainee response can be either 1) verbal acknowledgement, 2) behavioral change.

Context:
$\langle$ context string$\rangle$

Phrase:
$\langle$ phrase string$\rangle$

Format your response as follows. DO NOT DO ANY OTHER FORMATTING.:

'verbal acknowledgement': 'yes' if you predict the trainee to respond with a verbal acknowledgement otherwise 'no'

'behavioral change': 'yes' if you predict the trainee to respond with a behavioral change otherwise 'no'

Your output can be a combination of the two categories. For example:

\{
    'verbal acknowledgement': 'yes',
    'behavioral change': 'no'
\}

\section{GPT-4o Feedback Effectiveness Prediction From Human Annotations Prompt}\label{apd:fb_eff_hum_annots_prompt}

\textbf{System Prompt:}
You are an AI assistant specializing in predicting trainee responses during urology surgery training using the da Vinci robot. Your task is to analyze feedback from a trainer surgeon to a trainee surgeon, focusing on the trainee's reactions to feedback. 

You will categorize potential trainee responses into two types:

1. Verbal Acknowledgement: This includes any verbal or audible confirmation from the trainee indicating they have heard and understood the feedback. Examples include:
   - "Okay, I see"
   - "Uh-huh, got it"
   - "Understood"
   - "Yes, I'll do that"

2. Behavioral Change: This refers to any physical or observable adjustment made by the trainee that directly corresponds to the feedback received. For example:
   - If the trainer suggests tightening a suture, the trainee immediately pulls the suture thread more tightly.

Your role is to predict which type(s) of response the trainee is likely to give based on the specific feedback provided by the trainer. Consider the context of the surgical procedure and the nature of the feedback when making your predictions.

\noindent\textbf{User Prompt:}
Classify whether the following feedback phrase will lead to a trainee response, where a trainee response can be either 1) verbal acknowledgement, 2) behavioral change.

Feedback:
$\langle$ human annotation string$\rangle$

Format your response as follows. DO NOT DO ANY OTHER FORMATTING.:

'verbal acknowledgement': 'yes' if you predict the trainee to respond with a verbal acknowledgement, otherwise 'no'

'behavioral change': 'yes' if you predict the trainee to respond with a behavioral change otherwise 'no'

Your output can be a combination of the two categories. For example:

\{
    'verbal acknowledgement': 'yes',
    'behavioral change': 'no'
\}

\section{GPT-4o Feedback Component Classification From Auto Dialogue Prompt}\label{apd:fb_comp_prompt}

\textbf{System Prompt:}
You are an AI assistant specializing in classifying feedback during urology surgery training using the da Vinci robot. Your task is to analyze dialogue between a trainer and a trainee, focusing on categorizing the feedback into anatomic, procedural, and/or technical. The dialogue may contain multiple consecutive turns by the same speaker due to missed responses or varying speech patterns.

You will categorize the feedback into three types:

1. Anatomic: Familiarity with anatomic structures and landmarks. Examples include:
 - "Stay in the correct plane, between the 2 fascial layers."
 - "Avoid the blood vessels here."

2. Procedural: Pertains to the timing and sequence of surgical steps. Examples include:
 - "You need to suture this area first."
 - "You can switch to the left side now."

3. Technical: Performance of a discrete task with appropriate knowledge of factors including exposure, instruments, and traction. Examples include:
 - "Adjust the tension on the suture."
 - "Buzz it."

Your role is to predict which type(s) of feedback the phrase contains based on the specific feedback provided by the trainer. Consider the context of the surgical procedure and the nature of the feedback when making your predictions.

\noindent\textbf{User Prompt:}
Classify the feedback phrase into one or more of the following categories: 1) anatomic, 2) procedural, 3) technical. Do this while considering the given context of the last couple turns in the dialogue where the phrase is the last entry in the context.

Context:
$\langle$ context string $\rangle$

Phrase:
$\langle$ phrase string $\rangle$

Format your response as follows. DO NOT DO ANY OTHER FORMATTING.:

'anatomic': 'yes' if the feedback is anatomic otherwise 'no'

'procedural': 'yes' if the feedback is procedural otherwise 'no'

'technical': 'yes' if the feedback is technical otherwise 'no'

Your output can be a combination of the three categories. For example:

\{
'anatomic': 'yes',
'procedural': 'no',
'technical': 'yes',
\}

\section{GPT-4o Feedback Component Classification From Human Annotations Prompt}\label{apd:fb_clf_hum_annotprompt}

\textbf{System Prompt:}
You are an AI assistant specializing in classifying feedback during urology surgery training using the da Vinci robot. Your task is to analyze feedback from a trainer to a trainee, focusing on categorizing the feedback into anatomic, procedural, and/or technical.

You will categorize the feedback into three types:

1. Anatomic: Familiarity with anatomic structures and landmarks. Examples include:
 - "Stay in the correct plane, between the 2 fascial layers."
 - "Avoid the blood vessels here."

2. Procedural: Pertains to the timing and sequence of surgical steps. Examples include:
 - "You need to suture this area first."
 - "You can switch to the left side now."

3. Technical: Performance of a discrete task with appropriate knowledge of factors including exposure, instruments, and traction. Examples include:
 - "Adjust the tension on the suture."
 - "Buzz it."

Your role is to predict which type(s) of feedback the phrase contains based on the specific feedback provided by the trainer. Consider the context of the surgical procedure and the nature of the feedback when making your predictions.

\noindent\textbf{User Prompt:}
Classify the feedback phrase into one or more of the following categories: 1) anatomic, 2) procedural, 3) technical. Do this while considering the given context of the last couple of turns in the dialogue where the phrase is the last entry in the context.

Feedback:
$\langle$ human annotation string $\rangle$

Format your response as follows. DO NOT DO ANY OTHER FORMATTING.:

'anatomic': 'yes' if the feedback is anatomic otherwise 'no'

'procedural': 'yes' if the feedback is procedural otherwise 'no'

'technical': 'yes' if the feedback is technical otherwise 'no'

Your output can be a combination of the three categories. For example:

\{
'anatomic': 'yes',
'procedural': 'no',
'technical': 'yes',
\}

\section{Cosine Similarity Threshold Experiment}
\label{apd:cossim_thresh}

\begin{table}[H]
\centering
\begin{tabular}{p{1.5cm}|ccp{1.5cm}}
\textbf{Sim.  Thresh} & \textbf{Prec.} & \textbf{Recall} & \textbf{Prec.-Leaning Mean} \ \\
\hline

0 & 0.085 & 0.507 & 0.423 \\
0.1 & 0.077 & 0.823 & 0.566 \\
0.2 & 0.063 & 0.933 & \textbf{0.593} \\
0.3 & 0.053 & 0.947 & 0.579 \\
0.4 & 0.046 & 0.973 & 0.579 \\
0.5 & 0.044 & 1.000 & 0.588 \\

\hline

\end{tabular}
\label{tab:cossim_thresh_tuning}
\caption{Metrics for classifying segments as trivial hallucinations. Precision-Leaning Mean is calculated by $2\times\text{Precision} + \frac{\text{Recall}}{2}$.}
\end{table}

Trivial hallucinations are found using a method described in \cite{koenecke2024careless} where ASR is run twice on the same phrase and if the outputs are different then it is considered as a trivial hallucination. Having this "true label" hallucinations dataset, we apply the different cosine-similarity thresholds in the Hallucination Removal step that filters out hallucinations by thresholding the cosine similarity between audio segment and trainer and trainee voice samples. Note that we prioritize precision more than recall because a high recall does imply we pick up all hallucinations but it also means that all of those phrases do not get considered to have feedback since they would be classified as hallucinations. Further, the identification process almost always identifies empty phrases as unknown and hallucinations 

\section{Dialogue Reconstruction Feedback Detection Confusion Matrices}
\label{apd:fb_confusion_matrices}

\begin{table}[H]
\centering
\begin{tabular}{l|cc}
& \textbf{Pred: False} & \textbf{Pred: True} \\
\hline
\textbf{Label: False} & 2144 & 358 \\
\textbf{Label: True} & 474 & 613 \\
\hline
\end{tabular}
\label{tab:dialogue_cm}
\caption{Dialogue (off-the-shelf dialogue reconstruction) feedback detection confusion matrix.}
\end{table}

\begin{table}[H]
\centering
\begin{tabular}{l|cc}
& \textbf{Pred: False} & \textbf{Pred: True} \\
\hline
\textbf{Label: False} & 502 & 197 \\
\textbf{Label: True} & 180 & 500 \\
\hline
\end{tabular}
\label{tab:hall_removal_cm}
\caption{Hallucination Removal (our) dialogue refinement feedback detection confusion matrix.}
\end{table}

\begin{table}[H]
\centering
\begin{tabular}{l|cc}
& \textbf{Pred: False} & \textbf{Pred: True} \\
\hline
\textbf{Label: False} & 656 & 115 \\
\textbf{Label: True} & 115 & 493 \\
\hline
\end{tabular}
\label{tab:trainer_id_cm}
\caption{Trainer/Trainee ID (our) dialogue refinement feedback detection confusion matrix.}
\end{table}

\section{Example Cosine Similarities of Embeddings}
\label{apd:cos_sim_examples}
\begin{figure}[H]
    \begin{center}
        \includegraphics[width=0.5\textwidth]{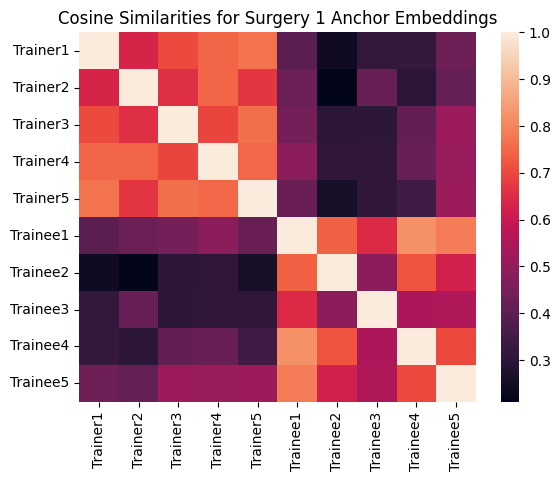}
        \vspace{-16.0pt}
        \caption{Cosine similarity between trainer and trainee anchor embeddings for surgery 1. Trainer1, ..., Trainer5 refer to the different audio examples for the same trainer and likewise for Trainee1, ..., Trainee 5. }
        \label{fig:cos_sim_examples}
    \end{center}
    \vspace{-20.0pt}
\end{figure}

\section{Detailed Temporal Event Detection}
\label{apd:detailed_temp_detection}
This approach relies on a moving fixed-length window of 10 seconds length with 5 sec overlap between the windows. These settings have been chosen empirically based on the observed average length of feedback. Note that feedback annotations in our dataset only have the beginning times, but not the duration. 

We classify each such moving window for the presence of feedback leveraging Audio, Text, and Audio+Text late fusion models. We apply \textit{Automated Speech Recognition (ASR)} with \textit{Whisper-1} \citep{radford2023robust} to obtain the text from a given timespan. For audio classification, we fine-tune Wav2Vec base model \citep{baevski2020wav2vec} with 95M parameters. 

For text classification, we fine-tune BERT base model with 110M
parameters \citep{devlin2018bert}. For multimodal classification, we apply a late fusion approach where we extract richer representations from the audio and text modalities as 256-dimension vectors. The representations are concatenated into one 512-dimension vector and passed via 2 fully-connected linear layers that reduce the dimensions to 64 and finally 2. This sequential architecture is augmented with ReLu activation and additional dropout in between. The additional steps can help the model calculate intermediate fusion features.

\end{document}